\documentclass[%
 reprint,
 amsmath,amssymb,
 aps,
]{revtex4-2}

\usepackage{graphicx}
\usepackage{dcolumn}
\usepackage{bm}

\usepackage[utf8]{inputenc}
\usepackage{amsmath}
\usepackage{amssymb}

\usepackage{amssymb}
\usepackage{scalerel}
\usepackage{hyperref}
\usepackage{cleveref}
\usepackage{verbatim}

\usepackage{lipsum}
\usepackage{amsfonts}
\usepackage{graphicx}
\usepackage{epstopdf}
\usepackage{algpseudocode}
\ifpdf
  \DeclareGraphicsExtensions{.eps,.pdf,.png,.jpg}
\else
  \DeclareGraphicsExtensions{.eps}
\fi

\usepackage{amsopn}

\begin{document}

\preprint{APS/123-QED}

\title{Numerical Computation of Effective Thermal Equilibria\\ in Stochastically Switching Langevin Systems}

\author{Benjamin L. Walker}
\author{Katherine Newhall}%
 \email{knewhall@unc.edu}
\affiliation{%
 University of North Carolina at Chapel Hill
}%

\date{\today}

\begin{abstract}
Stochastically switching force terms appear frequently in models of biological systems under the action of active agents such as proteins. The interaction of switching force and Brownian motion can create an ``effective thermal equilibrium'' even though the system does not obey a potential function. In order to extend the field of energy landscape analysis to understand stability and transitions in switching systems, we derive the quasipotential that defines this effective equilibrium for a general overdamped Langevin system with a force switching according to a continuous-time Markov chain process. Combined with the string method for computing most-probable transition paths, we apply our method to an idealized system and show the appearance of previously unreported numerical challenges. We present modifications to the algorithms to overcome these challenges, and show validity by demonstrating agreement between our computed quasipotential barrier and asymptotic Monte Carlo transition times in the system.
\end{abstract}

\maketitle


\section{Introduction}

Biological systems under the influence of microscale active agents such as proteins are frequently modeled using switching forces as the agents shift between different states~\cite{bressloff2017stochastic}. Examples include molecular motors~\cite{welte2004bidirectional}, crosslinked biopolymer networks~\cite{cao2019rheological}, and transient antibody crosslinking of antigens to mucus protein networks~\cite{jensen2019antibody,schroeder2020lps}. Protein action also plays a crucial role in organization of the DNA inside the cell nucleus, both in the form of stochastic crosslinking~\cite{alipour2012self, ea2015contribution, goloborodko2016compaction,schalbetter2017}, and more recently, protein loop extrusion~\cite{ganji2018real,he2020statistical,terakawa2017condensin}.  
Modeling these active agents in combination with passive diffusion leads to mathematical models with two sources of noise -- stochastically switching forces combined with stochastic Brownian motion.

In the case of crosslinking proteins,  prior work in the literature on modeling of the dynamic organization of DNA with polymer bead-spring models included stochastically-switching spring forces between beads representing 5kbp of DNA~\cite{hult2017enrichment,walker2019transient}.  These rapidly switching forces are on timescales faster than the time to reach thermal equilibrium, thus the system is in a constant state of disequilibrium.  However, in \cite{walker2019transient} we observed long-lived stable condensed clusters of beads consistent with experimental results, with the stochastic switching rate acting like an effective temperature.  Rapid switching produced low-temperature-like stable clusters, slow switching produced high-temperature-like amorphic arrangements, and intermediate switching times allowed for dynamic clusters with beads exchanging between clusters.

To explain the mechanism behind this emergent clustering behavior, we seek an effective thermal equilibrium.  Recall that 
if the forces in a system, $v(x)=-\nabla U(x)$, are the gradient of a potential function $U(x)$, the dynamics governed by the stochastic differential equation (SDE)
\begin{equation}
    \label{eqn:intro_langevin}
    dX = v(X) dt + \sqrt{2k_BT}dW
\end{equation}
approach, in the long time limit, a Boltzmann thermal equilibrium distribution given by
\begin{equation}
    p(x) \sim \exp\left(-\frac{U(x)}{k_B T}\right).
    \label{eqn:boltzmann}
\end{equation}
States $x$ that minimize $U(x)$ are long-lived stable configurations at temperatures small enough relative to the energy barriers of $U(x)$ separating such states, defined as the gap $\Delta U$ between the energy at the minimum and the energy at the lowest saddle point on the region of attraction of the minimum.
The mean transition time between the wells surrounding the energy-minimizing states can be computed asymptotically for vanishing temperature ($k_B T \to 0$): 
the time $\tau$ taken for the system to escape from a potential well under Brownian noise relates asymptotically to the energy barrier of the well following the Arrhenius equation given by
\begin{equation}
    \mathbb{E}\left[\log \tau\right] \sim \frac{\Delta U}{k_B T}.
\end{equation}
The most probably path (MPP) the system traverses as it makes one such transition can also be found asymptotically for vanishing temperature; it is a path that is everywhere parallel to the gradient of the energy landscape $U(x)$.

The study of energy landscapes and large deviations allows for valuable insights to be made by viewing systems through the lens of statistical thermodynamics.  In the cases of either stochastic switching of deterministic forces alone~\cite{bressloff2014path,newby2014spontaneous} or non-gradient forces with thermal noise~\cite{Ryter2012,cameron2012finding,Cameron2019},
 the derivation of a quasipotential in the small temperature limit successfully predict equilibrium distributions as well as transition times and paths.
We seek an effective equilibrium that takes into account both the thermal fluctuations and the stochasticity induced by the switching forces in the above-mentioned DNA bead-spring model.

A first thought to find an effective potential function $U(x)$ is
to simply time-average the force, thereby removing the switching. However, this approach significantly overestimates the effective strength of a strong force when the switching is not so fast. Considering a thought experiment of an infinitely high potential barrier that is only sometimes on, we can see that no matter the 
manner of the switching, the barrier remains infinitely high after averaging. However, as the particle could cross the barrier due to diffusion while it is off, transmission is clearly possible, contrary to the expectation of the naive time-averaging.

This time-averaging  takes the noise from the stochastic switching to zero first, and then considers the effects of thermal noise.  Rather, we seek a distinguished limit that takes the switching timescale to zero simultaneously with the thermal noise.  
To  simultaneously consider both sources of randomness while leveraging
the power of the energy landscape framework, we compute 
 a quasipotential $W(x)$ whose gradient represents an average force that 
generalizes the asymptotic properties of the potential function to non-gradient systems.
We build off the work in Ref.~\cite{newby2014spontaneous} that used a WKB approximation to construct a quasipotential for the Morris-Lecar equation, an ODE whose evolution depends on a stochastically-changing number of open ion gates.  
In our case, we consider both a switching state modeled by a continuous-time Markov chain and diffusive noise from Brownian motion.  We extend the WKB ansatz approach of \cite{newby2014spontaneous} to form the Hamilton-Jacobi equation for our problem.

The quasipotential $W(x;x_A)$ differs from a global potential function in that it is only defined in the basin of attraction of a particular fixed point $x_A$ of the deterministic system.  Since we are interested in using the quasipotential to predict expected transition times, we seek its values along the MPP the system escapes along.  As with MPPs for gradient systems, this path connects $x_A$ to a saddle point and is everywhere perpendicular to the gradient of the quasipotential.  To simultaneously find this path and the quasipotential along it we use the string method~\cite{weinan2002string} and its climbing variant~\cite{ren2013climbing}.  These string methods place a number of copies of the system, or images, along a path in phase space to form a `string'.  Each image is independently updated via gradient descent, then the images are reinterpolated along the path to keep them equally spaced in arclength.  Thus the path aligns itself with the gradient descent direction, converging to the MPP.  In the climbing variant, the final image ``climbs'' in energy in the direction tangent to the string, and gradient descends in all others, thus this final image converges to a saddle point, with the remaining images parameterizing the MPP from this saddle point.

In this work, we must couple the evolving string with finding the quasipotential.  Therefore we develop a numerical scheme to iteratively solve for the quasipotential along a path, then update the path based on the found quasipotential until it converges to the MPP.

While previous work generally looked at one- to three-dimensional problems~\cite{bressloff2017hamiltonian,brzezniak2015quasipotential,chen2019non,moore2015qpot,newby2014spontaneous,yang2019computing,zhou2012quasi}, we demonstrate an example for three particles in two dimensions for a total of six spatial dimensions. Under our proposed framework, several aspects of the problem present numerical challenges that were not reported in the previous literature. This numerical instability arises in both the implicit solver for the gradient of the quasipotential and the solver for the MPP. We find that Newton's method does not generally converge from the starting guesses we can make, and so we use a modified version of Newton's method with an additional fallback designed to ensure convergence. This also leads to high sensitivity in the Hessian matrix, which is used in the GMAM method for finding the MPP~\cite{heymann2008geometric,newby2014spontaneous}. This explains the choice of using the string method instead, which does not require the Hessian matrix. Still, we see significant high-frequency noise in the string method update, arising for larger numbers of images in the string. We present a set of numerical methods, along with code, that overcomes all of these challenges, and demonstrate 
validity by showing agreement between quasipotential barriers and Monte Carlo escape time asymptotics. 

In \cref{sec:formulation},  we derive the form of the Hamiltonian for our problem and describe numerical methods we apply to solve for the gradient of the quasipotential and most-probable transition paths. In \cref{sec:1d} we validate the formulation of the Hamiltonian on a 1d problem in which the string method is not necessary. In \cref{sec:2d} we then apply the full method to compute transition paths and asymptotic escape times for a system of three particles moving in two dimensions, revealing an important physical principle, that the interaction of stochastic switching and Brownian noise leads to a weaker effective force than would be expected by simple averaging. 
Finally, in \cref{sec:discussion} we review our contributions and note future directions of research.

\section{Formulation}
\label{sec:formulation}

In this section we detail the steps required to compute the gradient of the quasipotential $\nabla W$ along transition paths.
Full code to compute transition paths and quasipotential barriers for arbitrary problems of the form~\cref{eqn:langevin_main} and replicate all results in this work is available on Github~\footnote{\url{https://github.com/bwalker1/quasi-string-reprod}}.

In \cref{sec:hamilton_jacobi} we will derive the Hamiltonian and the associated Hamilton-Jacobi equation that define the gradient of the quasipotential $\nabla W$ for our problem.  In \cref{sec:dw_computation}, we will describe the algorithm that we use to address the additional numerical challenges in this version of the problem and solve the Hamilton-Jacobi equation for $\nabla W$. In \cref{sec:map} we describe the numerical procedure for finding most-probable transition paths which minimize the action, and in \cref{sec:meth_monte_carlo} we describe how we compute asymptotic escape times using Monte Carlo methods to validate the quasipotential values.

\subsection{Model Framework}
\label{sec:model_framework}
Taking \cref{eqn:intro_langevin} and introducing a  parameter $\epsilon = k_B T$, we modify the forces $v(x)$ to switch between different states to arrive at the general mathematical form of the equation for the dynamics.  
The configuration of the system is represented as the combination of a position $X_t \in \mathbb{R}^m$ and a switching state index $s_t \in \mathbb{Z}_1^n$, for some dimensionality $m$ and number of possible switching states $n$. The time evolution of the position 
follows an overdamped Langevin equation given by
\begin{equation}
    \label{eqn:langevin_main}
    dX_t = v(X_t; s_t) dt + \sqrt{2 \epsilon} dW
\end{equation}
where the switching state $s_t \in \{1,...,n\}$ affects the drift term, and the time evolution of the state $s_t$ follows a continuous-time Markov Chain (CTMC) process whose transition rate matrix $\frac{1}{\epsilon}S(X_t)$ depends on position. In this way the processes governing position and switching state are coupled. 
We chose the scaling of $\frac{1}{\epsilon}$ in front of $S$ to produce interaction between the CTMC switching noise and the $\epsilon=k_B T$ noise from Brownian motion at lowest order asymptotically, thus preparing to take a distinguished limit of small noise.

\subsection{Deriving the Hamilton-Jacobi Equation}
\label{sec:hamilton_jacobi}

The Hamilton-Jacobi equation that defines the quasipotential arises from making a WKB-like quasi-steady state assumption for the solution to a system of Fokker-Planck equations that are coupled by the transitions between  switching states.  We start by defining these coupled Fokker-Planck equations.
Let the force (drift) on the $i$-th position coordinate under switching configuration $s$ be represented as $v_i^s$. 
Recall the CTMC process transition matrix elements $S_{jk}$
hold the transition rate into state $j$ from state $k$.
Finally, let $p_s(x, t)$ represent the joint probability function between the discrete variable $s$ and the continuous position variables of each bead,
\begin{equation}
	 p_s(x, t) = \rho(x,t | s_t=s)\textrm{P}( s_t = s) \quad \textrm{for } s=1,2,\dots n
\end{equation}
where $\rho(x,t|s_t=s)$ is the conditional density for the process $X$ at time $t$ given that the force state is currently in state $s$.

Each individual $p_s$ will follow a Fokker-Planck equation associated with that state's drift term from \cref{eqn:langevin_main}, with an additional coupling term to represent transitions between states. These coupled Fokker-Planck equations take the form

\begin{equation}
    \frac{\partial p_s}{\partial t} = -\sum_{i=1}^m \frac{\partial}{\partial x_i}\left[v^s_i p_s\right] + \epsilon \sum_{i=1}^m \frac{\partial^2}{\partial x_i^2} \left[ p_s \right] + \frac{1}{\epsilon} \sum_{k=1}^n S_{sk} p_k
\end{equation}
with the associated steady state equation,
\begin{equation}
    \hspace{0.45cm}0 = -\sum_{i=1}^m \frac{\partial}{\partial x_i}\left[v^s_i p_s\right] + \epsilon \sum_{i=1}^m \frac{\partial^2}{\partial x_i^2} \left[ p_s \right] + \frac{1}{\epsilon} \sum_{k=1}^n S_{sk} p_k
    \label{eqn:steady_state}
\end{equation} for $s=1,2,\dots n$.

If the drift term $v$ did not depend on the switching process $s_t$, the Langevin process for $X_t$ would have a potential function constructed from a path integral of $v$.
However, in our formulation in which the drift function $v$ exhibits random switching, it is no longer the gradient of a potential function, and so such a $U$ cannot be found.  Intuitively, in the small-noise limit $\epsilon\to 0$, in which the magnitude of the Brownian noise goes to $0$ as the rate of the stochastic switching  goes to infinity, there is no diffusion and the forces $v$ exist in a superposition of the $v^s$ according to the steady state distribution of $s$.  

The above informs our choice of the WKB-like ansatz for the steady-state distribution $p_s(x)$ in the form
\begin{equation}
	\label{eqn:wkb_ansatz}
	p_s(x) = r_s(x) \exp\left( -\frac{1}{\epsilon} W(x)  \right)
\end{equation}
for $s=1\dots n$ as $\epsilon\to 0$, similar to the one employed in~\cite{newby2014spontaneous} for purely stochastically-switching forces (no diffusion).
 We see here that $W(x)$ takes the place of the potential $U(x)$ in \cref{eqn:boltzmann}; $W(x)$ is the quasipotential.
 The pre-exponential term $r_s$ superimposes the different states $s$.
 
We then plug \cref{eqn:wkb_ansatz} into \cref{eqn:steady_state}, seeking equations for $r_s(x)$ and $W(x)$ given by the order $\frac{1}{\epsilon}$ terms, which are the lowest order in $\epsilon$.  In this way, \cref{eqn:wkb_ansatz} differs from a typical WKB expansion in which there would be no pre-exponential term present at lowest order. This $r_s$ term only encapsulates the relationship between states -- the full pre-exponential term would emerge at higher order in $\epsilon$.

The resulting order $\frac{1}{\epsilon}$ equation has the form
\begin{equation}
\label{eqn:Mr0}
M(x, \nabla W) r(x) = {0}
\end{equation}
where the matrix $M$ depends on both the position $x$ and the gradient of the quasipotential $\nabla W$.  The 
 vector ${r}(x) = (r_1,r_2,\dots,r_n)^T$ has  components  $r_s(x)$ for each state $s$.
Details are shown in \cref{sec:hamiltonian_derivation}. The $n\times n$ matrix $M$ is given by
\begin{equation}
M(x, \nabla W) = D(\nabla W) + A(x, \nabla W) + S(x),
\end{equation}
 the sum of three matrices corresponding to the three terms in \cref{eqn:steady_state}: $D$, the diffusion matrix; $A$, the advection matrix; and $S$, the switching matrix.
Note that $S$ is unchanged from its original definition as the CTMC transition rate matrix and serves the purpose of coupling the different states, while $D$ and $A$ are diagonal matrices with diagonal elements for each switching state $s=1\dots n$ given by
\begin{equation}\begin{aligned}\label{eq:DandA}
    D_{ss} = \sum_{i=1}^m \left( \frac{\partial W}{\partial x_i}\right)^2 \quad\textrm{and} \quad
    A_{ss}  = \sum_{i=1}^m v_i^s \frac{\partial W}{\partial x_i}.
\end{aligned}\end{equation}

We can observe that in order to have non-trivial solutions to the system \eqref{eqn:Mr0} we must have that $\det M(x, \nabla W) = 0$. Choosing the Hamiltonian as the greatest eigenvalue of $M$, $\mathcal{H}(x, p) = \textrm{max}\; \lambda \textrm{ s.t. } M(x,p) u = \lambda u$, 
 we see that having non-trivial solutions to the system \eqref{eqn:Mr0} is equivalent to 
 the Hamilton-Jacobi equation, $\mathcal{H}(x,p) = 0$.  Note that we have introduced a new variable $p$ as the second argument, by analogy to the typical form of the Hamilton-Jacobi equation. This means that our solution for $\nabla W$ will be given by the value of $p$ that solves the Hamilton-Jacobi equation.

Combining the above with an additional curl-zero constraint, we obtain the pair of equations 
\begin{align}
	\label{eqn:cond1}
	\mathcal{H}(x, \nabla W(x)) &= 0 \\
	\nabla \times \nabla W(x) &= 0
\end{align}
that uniquely define the gradient of the quasipotential.
In this work we will compute the quasipotential along most-probable paths (MPPs) parameterized as $\phi(s) : [0, 1] \to \mathbb{R}^m$. On the interior of the path $0<s<1$, \cref{eqn:cond1} will define a convex surface, and the curl constraint that defines the unique solution can be replaced with a constraint evaluated only on the path:
\begin{equation}
\label{eqn:cond2}
\left.\nabla_p \mathcal{H}(x,p)\right|_{p=\nabla W(x)} \parallel \frac{d\phi}{ds}.
\end{equation}
We will require that the endpoints of this path are fixed points of the ``deterministic dynamics,'' defined by
the limit $\epsilon \to 0$ of \cref{eqn:langevin_main}.  Recall that in this limit, the Brownian noise term vanishes, and the CTMC switching rates go to infinity such that the system always exists in a superposition of states consistent with the steady state of the switching matrix $S({x})$ at its current position. These dynamics can be expressed as 
\begin{align}\label{eq:deterministic}
\begin{split}
    \frac{d{x_i}}{dt} &= \sum_{k=1}^n v_i^k r_k \; \textrm{for } i=1\dots m \\
	{r} &= \textrm{null } S \qquad \sum_{k=1}^n r_k = 1
\end{split}
\end{align}
where $v_i^s$ defines the force on coordinate $i$ when in switching state $s$, and ${r}$ is a null-vector of $S({x})$ appropriately normalized so that it represents the steady-state distribution of the CTMC at fixed position $x$.
While $\mathcal{H}(x, 0)=0$ everywhere, we show in \cref{sec:det_fp} that at fixed points of the deterministic dynamics, $\mathcal{H}(x,0)=0$ is a minimum of $\mathcal{H}$, and therefore the unique solution (note that $\mathcal{H}$ is a convex function). Thus, at the endpoints we necessarily have $\nabla W = 0$, and on the interior $\nabla W$ is defined by the simultaneous solution of \cref{eqn:cond1,eqn:cond2}.
  
Solving \cref{eqn:cond2} for the MPP $\phi$ can be done using variational methods such as the string method \cite{weinan2002string}. However, in the non-gradient case, as the quasipotential can be thought of as propagating along caustics, we do not know the true value of the quasipotential except on an MPP. This creates the need  for an iterative algorithm that alternates between estimating $\nabla W$ using the current value of $\phi$ and using the computed values of $\nabla W$ to update the path $\phi$.  We discuss this procedure in more detail next.

\subsection{Solving for $\nabla W$}
\label{sec:dw_computation}
We will now describe the numerical optimization procedure by which we obtain $\nabla W$, the gradient of the quasipotential, through which we obtain the final quasipotential by numerical integration. 
This will require formulating the simultaneous solution of \cref{eqn:cond1,eqn:cond2} as an optimization problem using Lagrange multipliers and then plugging in the Hamiltonian constructed in \cref{sec:hamilton_jacobi}. 

Our approach for computing $\nabla W$ is based on the Newton's method equations in \cite{newby2014spontaneous}, which are presented for an arbitrary Hamiltonian $\mathcal{H}$, and also require the gradient and Hessian of $\mathcal{H}$ with respect to the momentum variables $p$ (see \cref{eq:piterate,eq:lambiterate}). Using the Hamiltonian derived in \cref{sec:hamilton_jacobi} along with the differentiation formulas shown in \cref{sec:eig}, we are able to apply these equations to our problem.

In practice, we have observed that direct application of these Newton's method equations often fail to converge for the problems we have attempted. To address this, we have added a fallback scheme that is designed to
 make guaranteed iterative improvements until the region in which Newton's method converges is reached.

This fallback method reframes \cref{eqn:cond1,eqn:cond2} as the problem of maximizing the dot product with $\frac{d\phi}{ds}$ on the convex surface $\mathcal{H}(x,p)=0$.  The fallback method moves the current guess a small distance in the direction of $\frac{d\phi}{ds}$ projected onto the normal of the surface $\mathcal{H}(x,p)=0$, and then applies Newton's method to return to a solution of \cref{eqn:cond1}, $\mathcal{H}(x,p)=0$. Note that the application of Newton's method in the fallback is only on \cref{eqn:cond1}, as opposed to the outer Newton's method which seeks a simultaneous solution to \cref{eqn:cond1,eqn:cond2}.

If the outer Newton's step fails to improve the quality of the solution, this fallback is used instead, which is guaranteed to produce a better solution. In practice, the fallback is able to quickly bring the guess close enough to the true solution for Newton's method to begin to converge quadratically. Additional details of this routine are written in \cref{sec:app_quasi_solver}.

\subsection{Computing Minimum Action Paths}
\label{sec:map}
Previous work on computing quasipotentials has either computed the quasipotential along MPPs~(for example, \cite{bressloff2014path,newby2014spontaneous}) or on a grid using an upwind scheme~\cite{newby2014spontaneous,yang2019computing}. As we are interested in transition asymptotics, we restrict this work to considering computing quasipotentials along MPPs, and so the routine for computing the quasipotential must simultaneously search for a MPP.

MPPs are computed by representing a path connecting two points $x_a, x_b$ in the position space of the system using a path $\phi(s)$ such that $\phi(0) = x_a$, $\phi(1) = x_b$. We discretize the path $\phi$ as a sequence of ``images'' representing states along the transition path: $\phi_1, \phi_2, \phi_3, ..., \phi_N$. These images are chosen so that the arc length is constant between images, e.g. $||\phi_{k}-\phi_{k-1}||=\textrm{const}$. To compute the MPP, we make an initial guess of the transition path (typically linearly interpolated) and then apply the string method~\cite{weinan2002string}, in which each image along the path is moved a small distance opposite the direction of $\nabla W$. Compared to typical applications of the string method, there is an additional step of applying the implicit solver to find the value of $\nabla W$ at each image, which itself depends on the current direction of the string.

Because we are looking for escape trajectories from a well, we allow the final image of the string to move according to the climbing string method~\cite{ren2013climbing}. The final image moves in a different direction: letting $\nabla W_f$ refer to the gradient of the quasipotential at the final image, and $d\phi_f$  refer to the direction of the string at the final image, the direction is given by
\begin{equation}
	\nabla W_f - (1+\alpha) \textrm{proj}_{d\phi_f} \nabla W_f
\end{equation}
for some $\alpha>0$.
This can be interpreted as descending in the directions orthogonal to $d\phi_f$, but climbing in the direction parallel to $d\phi_f$, as the projection term inverts that component.

Details of the algorithm are presented in \cref{sec:app_string}.

\section{1-dimensional case}
\label{sec:1d}
To demonstrate our formulation of the Hamiltonian~(\cref{sec:hamiltonian_derivation}) and correspondingly the quasipotential, we begin by applying our method to the one-dimensional case of a single bead subject to a constant excluded volume force and an on-off switching attractive force pulling it towards  
the origin. We note that the presence of only a single spatial dimension means the transition path is trivially known; this frees us from the additional optimization step of computing the most-probable transition path out of the minimum.

In this case, there are only two states of the switching force corresponding to whether the attractive force is switched off or on -- we label these forces $v^1$ and $v^2$, respectively, in line with the notation introduced in \cref{sec:hamilton_jacobi}.  They are given by
\begin{align*}
 v^1(x) &= a_{ev} x \exp\left( -\frac{x^2}{c_{ev}} \right) \\
 v^2(x) &=  a_{ev} x \exp\left( -\frac{x^2}{c_{ev}} \right) - k  x.
\end{align*}
In this example, we use parameters $k = 5$, $a_{ev} = 3$, $c_{ev} = 0.5$. We note that the method is independent to the choice of these parameters, and the choice is simply motivated to create a well with a basin of attraction extending to $|x| \approx 1$. 

Correspondingly, the transition rate matrix takes the form
\begin{equation}
\label{eqn:1d_S}
    \frac{1}{\epsilon} S = \frac{1}{\epsilon} \begin{pmatrix}
    -a(x) & c \\
    a(x) & -c
    \end{pmatrix}
\end{equation}
with $a(x)$ being some decreasing function of distance from the origin, and $c$ a constant that is here taken to be $0.5$.  Note the $1/\epsilon$ scaling of the CTMC to couple the switching rates with the magnitude of the Brownian motion, thus allowing us to take the limit as both fluctuations go to zero with $\epsilon\to0$.

We consider three different choices of $a(x)$, chosen to create quasipotential barriers of different heights, to allow us to show agreement in three different cases.  These functions $a(x)$ are
\begin{align}
    a_1(x) &= 2 e^{-3|x|^2} \\
    a_2(x) &= \frac{2}{1+e^{20(|x|-0.75)}} \label{eq:affinity}\\
    a_3(x) &= \frac{4}{1+e^{20(|x|-0.75)}}.
\end{align}

Because the most-probable transition path can only move in the single dimension of the problem, we simply compute the quasipotential along some interval $(0, x_0)$ such that $x_0$ is to the right of the ``saddle point'' of the quasipotential (which in one dimension is in fact a maximum). For the above choices, the maximum is in the vicinity of $x=1$, so we compute values through $x_0=2$.

In order to validate the size of the computed quasipotential barrier, we compare to escape times, computed as described in \cref{sec:meth_monte_carlo}, using the modified Euler-Maruyama method. Note that the transition rate from state $s=1$ to $s=2$ depends on the changing variable $x$ and so waiting times are resampled each timestep; the rate $c$ is constant and so the waiting time from state $s=2$ to $s=1$ can be preserved until it is reached.

\begin{figure}
    \centering
    \includegraphics[width=\linewidth]{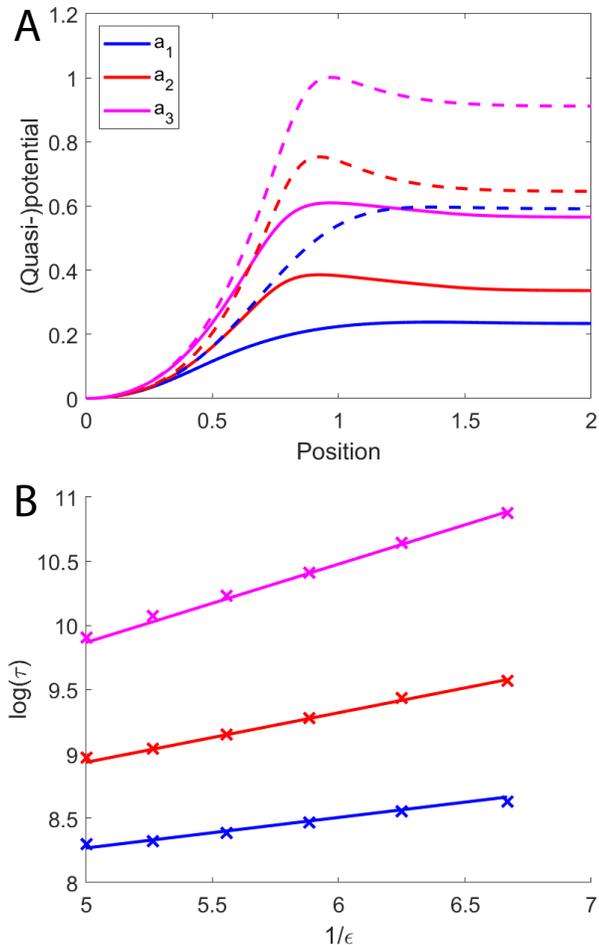}
    \caption{(Color Online) Comparison of Quasipotential and Escape Time Asymptotics for three different affinity functions. (A) Comparison of quasipotential along the string with the deterministic average, illustrating that the deterministic average would not agree with Monte Carlo statistics. (B) Average Monte Carlo escape times (points) showing  linear behavior whose slope is predicted by the quasipotential string.}
    \label{fig:res_1d}
\end{figure}

\Cref{fig:res_1d}(A) shows the results of computing the quasipotential for each of the three affinity functions, along with the deterministic energy computed by numerically integrating the deterministic force from \cref{eq:deterministic}. This shows that while the two agree on the location of the highest point of the barrier, they disagree significantly on its height. \Cref{fig:res_1d}(B) shows the average Monte Carlo escape times for different values of $\epsilon$, compared to lines whose slope is given by the height of the quasipotential barrier. The quasipotential barrier height clearly predicts well the slope followed by the escape times, confirming that our formulation of the quasipotential is consistent with the theory and preparing us to add the string descent element in higher-dimensional problems.

We also recall the thought experiment of the infinitely high switching barrier from the introduction. \Cref{fig:res_1d} confirms that the interaction between Brownian noise and stochastic switching leads the deterministic average to significantly overestimate the effective barrier, in this case by a factor of roughly two. This reinforces the need to properly consider the distinguished asymptotic limit in both sources of noise.

\section{2-dimensional case}
\label{sec:2d}
In this section, we detail the model of three beads moving in two dimensions that exhibits the same mixing behavior that originally inspired this work. We characterize the four stable states of the system, and show that the system spends most of its time in those states in which a pair of the beads are bound together. We compute a most-probable escape path from one of these states, and show that it agrees with Monte Carlo simulated escape times for transitions between bound states.

\subsection{Model}
\label{sec:2d_model}
Our interest in stochastically switching Langevin systems is originally motivated by prior work on modeling dynamics of chromosomes inside a yeast nucleus~\cite{walker2019transient}. We consider here a toy model using the same functional forms of forces, while noting that the methods apply in general to any type of stochastically switching force between particles.

We consider three beads, affected by three forces: bonding, repulsion (excluded volume) and a global confinement force. These are collectively sufficient to produce multiple stable equilibria in the system.
These forces have the forms:\\
Confinement:
\begin{equation}
    f_{\textrm{c}}^i(x_i) = - \eta x_i
\end{equation}
Excluded Volume:
\begin{equation}
    f_{\textrm{EV}}^i(\{x_j\}_j) = \sum_{j \neq i} a_{ev} \left(x_i-x_j\right) \exp\left(-\frac{\left(x_i-x_j\right)^2}{c_{ev}}\right)
\end{equation}
Attraction:
\begin{equation}
    f_{\textrm{bond}}^i(\{x_j\}_j) = \sum_{j \neq i} k b_{ij} \left(x_j-x_i\right)
\end{equation}
where $b_{ij}=1$ if beads $i$ and $j$ are presently bonded and zero otherwise,
and we use parameter values $k = 5$, $a_{ev} = 2$, $c_{ev} = 0.5$, $\eta=1$.  With this notation, the entries $b_{ij}$ stochastically switch as beads bind and unbind.  Thus, only the binding force $f_{\textrm{bond}}$ is stochastic; the others are deterministic.

This gives the following SDE for the position of bead $i$:
\begin{equation}
    \label{eqn:main_sde}
    dX_i = \left( f_c^i + f_{\textrm{EV}}^i + f_{\textrm{bond}}^i \right) dt + \sqrt{2 \epsilon} dW
\end{equation}
where $\epsilon$ is a small positive parameter that controls the amount of stochasticity in the system, and also appears in the switching, as described below.

The stochastic switching of the bonding term models the crosslinking proteins that bind two nearby beads.  
Each bead can be either unbound, or bound to a single other bead. Bonds are symmetric. If a bond is formed between two beads, assume that the lifetime of the bond is an exponentially distributed random variable with rate $c$, meaning it has an expected lifetime of $\frac{1}{c}$. To simulate it, one can simply draw such an exponentially distributed random variable and use it as the lifetime.

It is natural to think that crosslinking proteins would be more likely to bind beads that are closer together.  Therefore, by analogy to the form of \cref{eqn:1d_S} we include an ``affinity function'' $a(r)$ dependent on pairwise distances for the binding rate.
Specifically, $a(r)$ gives the (exponential process) rate at which a bond forms between two currently unbound beads $i, j$ with positions ${x}_i, {x}_j$ that are separated by a distance $r=|{x}_i-{x}_j|$. 
We note that this rate is only meaningful until a bond forms -- a rate of $a(r)$ is equivalent to stating that in an infinitesimally short time $dt$, there is a probability $\frac{dt}{a(r)}$ that a bond forms. However, note that as the system moves in time, these probabilities will change accordingly.
In this section, we will use $a=a_2$, given in \cref{eq:affinity}, which is 
\begin{equation*}
    a(x) = \frac{2}{1+e^{20(|x|-0.75)}}.
\end{equation*}

Following the framework of \cref{sec:model_framework}, the CTMC switches between the different binding configurations.  These binding configurations corresponding to the states of the Markov chain are enumerated as: all three beads unbound ($s=1$), bead 1 bound to 2 ($s=2$),  bead 1 bound to 3 ($s=3$),  bead 2 bound to 3 ($s=4$).
The corresponding transition rate matrix $S$ takes the form
\begin{equation}
    \frac{1}{\epsilon}S = \frac{1}{\epsilon}\begin{pmatrix}
    b & c & c & c \\
    a(x_1-x_2) & -c & 0 & 0 \\
    a(x_1-x_3) & 0 & -c & 0 \\
    a(x_2-x_3) & 0 & 0 & -c
    \end{pmatrix}
\end{equation}
with $b = -a(x_1-x_2)-a(x_1-x_3) -a(x_2-x_3)$. Recall again the fixed rate $c=0.5$ describes bonds breaking.

\begin{figure*}
    \centering
    \includegraphics[width=0.99\textwidth]{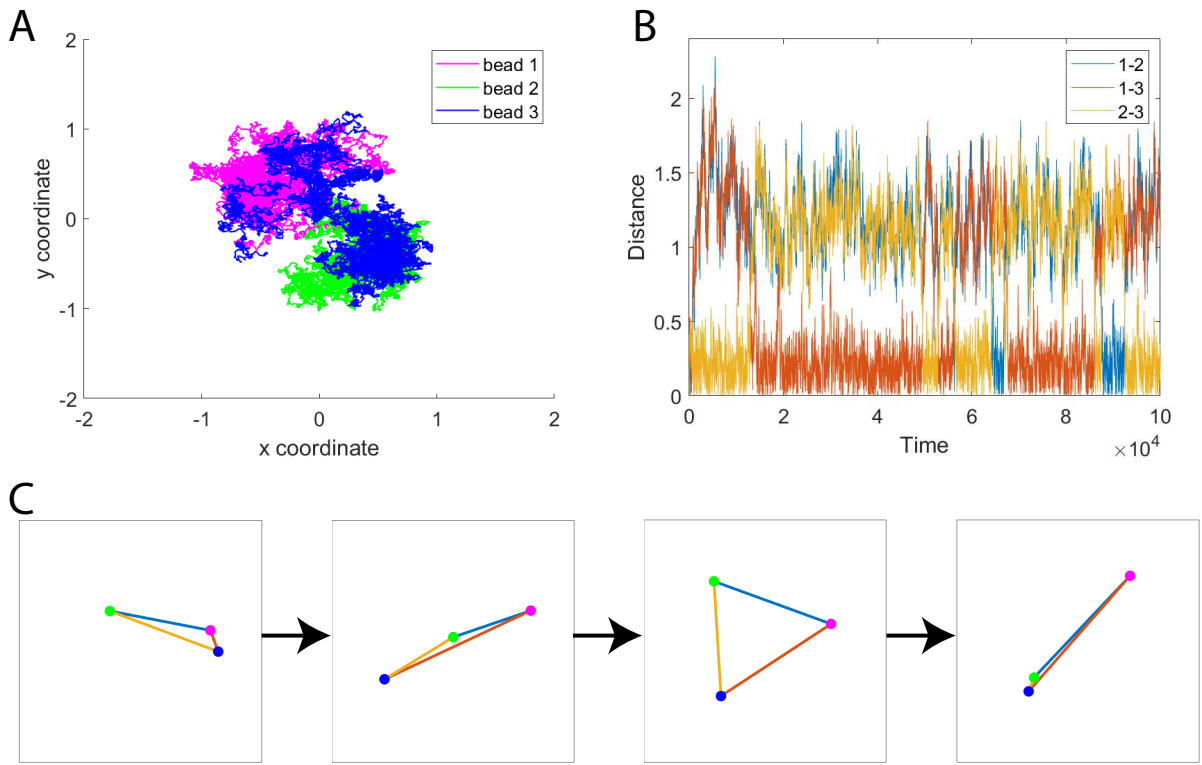}
    \caption{(Color Online) (A) Trajectories taken by three beads over the first 20000 timesteps of a simulation. (B) Plot of pairwise distances between the three beads over time. Observe that at any time, one distance is small (between the two currently bound beads) and the other two distances are large, with which pair is bound rapidly switching. (C) Illustration of a transition between two bound states. The bound beads separate, arranging into a line. The line then morphs into a triangle. Finally, two beads again approach and enter a bound state. }
    \label{fig:3bead_model}
\end{figure*}

To illustrate the qualitative behavior of this constructed model system, \cref{fig:3bead_model} demonstrates a sample simulated trajectory. In particular, we can see that this model replicates the ``mixing" property that motivated this research, with rapid switching between which pair of beads (analagous to a cluster) is currently bound. This motivates the further investigation into the behavior of this system, and the stability of the bound state.

\subsection{Computing Most-Probable Escape Paths}
\label{sec:res_mpp}

We now proceed to demonstrating the process by which we compute a most-probable transition path out of a basin of attraction using the quasipotential climbing string method.

\begin{figure}
    \centering
    \includegraphics[width=0.99\linewidth]{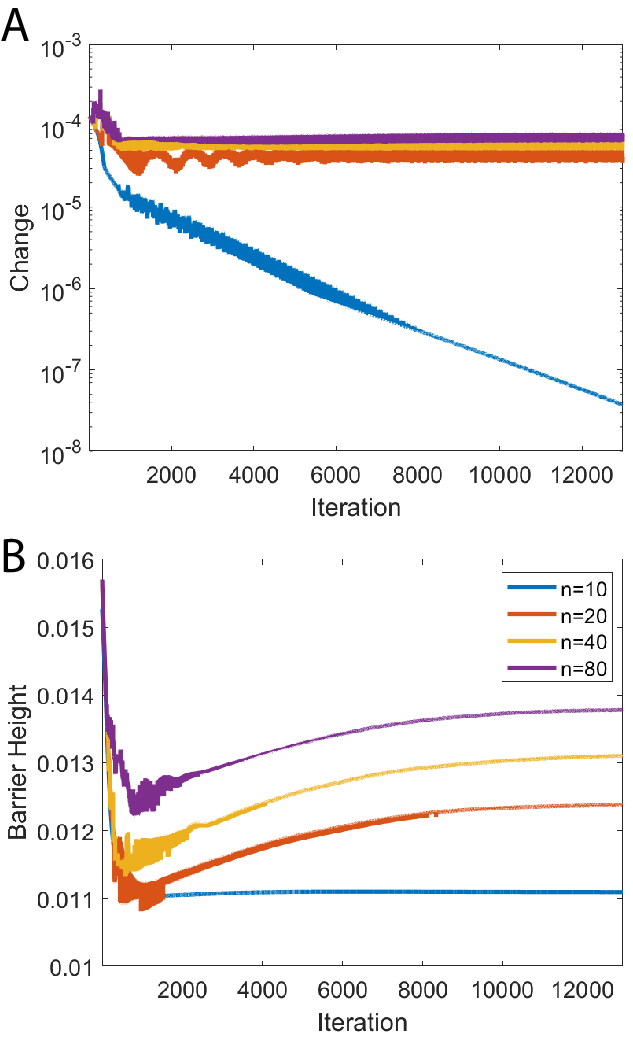}
    \caption{(Color Online) (A) Maximum change over all images from previous iteration, showing convergence only in the case of 10 images. (B) Quasipotential barrier height over iterations, showing that the $n=10$ converges to a barrier height of approximately $0.011$.}
    \label{fig:string_err}
\end{figure}

Compared to previous work on string descent, our problem exhibits a greater level of numerical instability. We begin by initializing a climbing string to search for an escape path out of the bound state minimum. Based on preliminary observations that emergence of high-frequency error in images along the string prevents convergence, we consider the most natural way to reduce such error, which is by reducing the number of images. In \cref{fig:string_err}, we plot the total change over all images since the previous iteration  (A), and the height of the quasipotential barrier (B), over the course of iteration. We observe that while a string with $10$ images exhibits a decline in change indicating convergence, strings with more images do not, with the change instead settling into a high-frequency oscillation.
 
\begin{figure}
    \centering
    \includegraphics[width=0.99\linewidth]{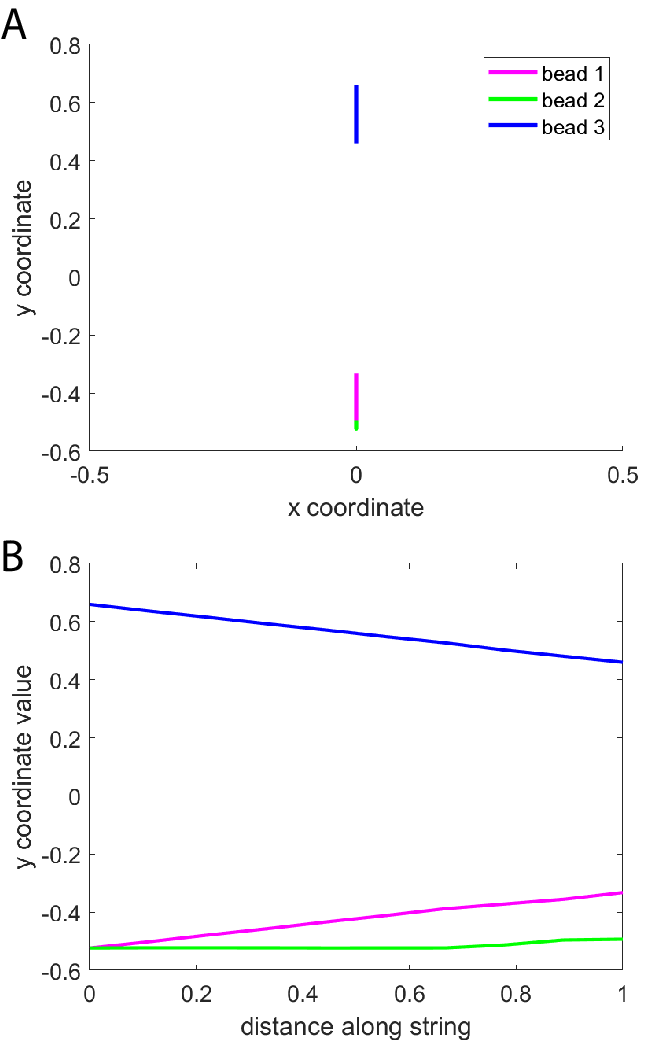}
    \caption{(Color Online) Visualization of escape path computed using quasipotential string descent, connecting bound state (\cref{fig:3bead_model}C) to saddle leading to line (\cref{fig:3bead_model}D). Note that this transition happens entirely on the y-axis. (A) Trajectory shown in 2 dimensions. (B) Plot of just the y-coordinate varying along the string. Note that the behavior is not simply linear, especially visible in the case of bead 2 (green).}
    \label{fig:string_main}
\end{figure}

Based on these results, we will use our string with $10$ images to determine the asymptotic transition time between bound states. This path is shown in additional detail \cref{fig:string_main}. Panel A shows the two-dimensional trajectory, which lies entirely along the y-axis, consistent with the observation in \cref{fig:3bead_model}(C,D,E) in which the transition passes through a line. Panel B shows the values of the coordinates along the string, demonstrating that they do not change in a direct linear fashion. With this string, we can now compute a quasipotential barrier and compare to the Monte Carlo asymptotics.

\subsection{Monte Carlo escape statistics}
\label{sec:res_string_mc}
To validate our quasipotential string descent method, we use it to predict escape time asymptotics from the bound state. As mentioned in \cref{sec:2d_model}, our test system exhibits switching between various permutations of the stable bound state, spending only minimal time in other stable configurations. For this reason, we compute escape times by initializing the simulation with beads $1$ and $2$ bound together, and measure time until a different pair of beads becomes bound, by a criterion that the newly bound beads should be separated by a distance under $0.3$ and the original bound pair is separated by a distance of at least $1$. While this criterion represents entering a new bound state, as opposed to simply reaching the edge of the basin of attraction, these two events take the same amount of time asymptotically to the lowest (logarithmic order). Due to the multidimensionality of the system combined with the stochastic switching, directly determining the edge of the basin of attraction is not trivial.

\begin{figure}
    \centering
    \includegraphics[width=0.99\linewidth]{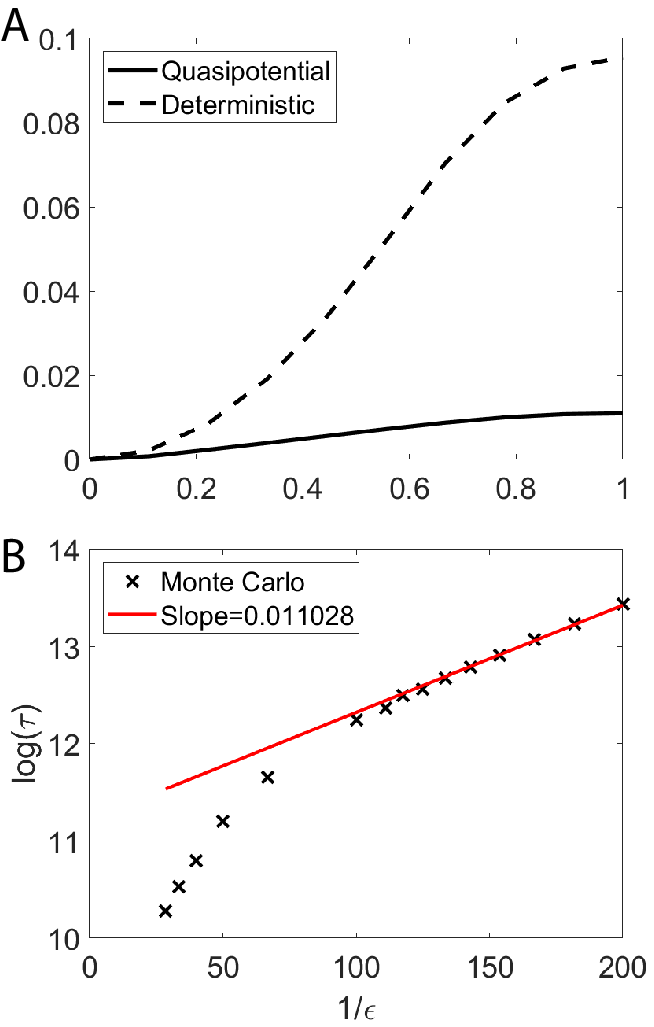}
    \caption{(Color Online) (A) Quasipotential (solid) and deterministic energy (dotted) along transition path from \cref{fig:string_main}. Quasipotential barrier height is approximately $0.011$. (B) Comparison of asymptotic escape times computed via Monte Carlo simulation to slope taken from quasipotential barrier height.}
    \label{fig:string_mc}
\end{figure}

We run this simulations for a collection of values of $\epsilon \in [0.05, 0.35]$, and compute mean escape time $\hat{\mu}$ as described in \cref{sec:meth_monte_carlo}. In \cref{fig:string_mc}(A), we see that for values of $1/\epsilon \geq 100$ ($\epsilon < 0.01$) the escape times are linear in the log-log plot, as predicted by the asymptotic relation, with a slope of approximately $0.118$. We then compare this to the quasipotential barrier computed by integrating $\nabla W$ along the most-probable escape path computed in \cref{sec:res_mpp}, which is shown in \cref{fig:string_mc}(B) to be approximately 0.11. This lines up quite closely with the value computed from Monte Carlo statistics. On the other hand, the value computed using the deterministic average, equal to approximately $0.96$, does not.

\section{Discussion}
\label{sec:discussion}
In this paper, we extended the theory of quasipotentials to systems of Langevin equations 
with an additional source of stochasticity arising from a continuous-time Markov process switching the force term, a natural mathematical framework for modeling systems in biology under the effect of protein binding/unbinding mechanics.
We demonstrated that under the interaction of switching forces and Brownian motion, an effective potential is generated that is weaker than predicted by a simple time averaging of the switching force. We further demonstrated that the quasipotential represents an effective thermal equilibrium in the low-noise regime, and can be used to predict asymptotic transition times between metastable states created by such switching forces. This sets the stage for further analysis of switching models of biological systems, representing states created by thermal action of proteins as minima in an effective energy landscape given by the quasipotential and transitions as most-probable paths in this landscape, only made possible by understanding the interaction between both sources of stochasticity.

By taking a distinguished limit simultaneously in both sources of noise, we derived the Hamiltonian for this problem
and demonstrated the numerical problems that arise when using the standard methods in the literature.
We developed modifications to these methods allowing us to compute most-probable transition paths and quasipotential barriers along these paths.
These quasipotential barriers accurately predict escape times, reinforcing that this simultaneous treatment of both noise sources is required over a time average of only the stochastic switching to produce a single deterministic force before considering the effects of thermal noise. 
This deterministic average significantly overestimates the effective energy barrier, and thus the stability of a pseudo-equilibrium state such as the gene clusters observed in \cite{walker2019transient}.
Thus we have shown that an effective thermal equilibrium can be constructed by simultaneously incorporating both the stochastic switching that pushes the system out of equilibrium and the thermal noise.

The inclusion of Brownian motion in our framework contrasts with previous work in which the only source of noise come from random switching between deterministic differential equations~\cite{bressloff2017hamiltonian,newby2014spontaneous}.
With the correct choice of scaling between the thermal noise and the stochastic switching rates, we were able to 
demonstrate that a similar WKB ansatz approach can be applied to derive the Hamilton-Jacobi equation for this framework.  

Our example 6 dimensional system produced new numerical issues
not reported by others computing quasipotentials for 1 to 3 dimensional systems in previous literature~\cite{bressloff2017hamiltonian,brzezniak2015quasipotential,chen2019non,moore2015qpot,newby2014spontaneous,yang2019computing,zhou2012quasi}.
These numerical issues arose in two separate places in our algorithm: the implicit solve step for $\nabla W$, as discussed in \cref{sec:dw_computation}, and the optimization method for finding the most probable transition path, as discussed in \cref{sec:res_mpp}. In each case, we showed how to modify the algorithm to overcome these issues. As solving for $\nabla W$ is a convex optimization problem, it is expected that there should exist a good solver, and our modified Newton's method seems to fit this role, with guaranteed convergence under a short number of iterations from practical considerations. 
However, there is room for future work in further understanding the error in the MPP computation step, as the current remedy of reducing the number of images leaves desired the ability to compute paths on a finer resolution. We reiterate that the most commonly observed alternative to the string method, the Geometric Minimum Action Method (GMAM), does not alleviate this problems as it runs into instability due to high sensitivity in the Hessian matrix. However, it is conceivable that continued work on the method used for this optimization step could produce a more stable algorithm in higher numbers of images.

Applying our methods to our idealized model, we showed how a stochastically switching pairwise crosslinking force could create effective bound states, with the system exhibiting switching between which pair of beads is in the effective bound state. We note that this is not simply switching between which pair of beads are currently bound by the switching crosslinking force - when beads are in this so-called effective bound state, they are switching back and forth between being bound and unbound, but on average are bound enough of the time to remain in close proximity until a rare event through a combination of Brownian noise and switching of the force allows them to separate and a new pair of beads to enter the effective bound state. We showed the ability to accurately predict the asymptotic timescales on which this escape occurs, showing strong agreement between the slope of Monte Carlo simulations of the timescale and the height of the quasipotential barrier, in line with the Arrhenius law of escape times. This validates our methods and code and sets the stage for use in further applications.

Our work here was originally motivated by observations in \cite{walker2019transient} in which we observed particles representing beads in a polymer model of the yeast genome associating into clusters as though on an effective energy landscape, even though the stochastically switching forces meant the system was never truly in equilibrium. In this work, we were able to mathematically demonstrate that the observed behavior can be explained by an effective energy landscape through analysis of a reduced model. Having derived the Hamilton for a general overdamped Langevin system and the numerical methods to compute transition paths, 
we would like to apply our methods in future work to analyze the clustering states observed in this model as fixed points of the associated deterministic dynamics, and then compute the quasipotential barrier to escape from these states to understand the stability of each clustering state in terms of system parameters. 

One of the challenges facing scaling up this method to more particles is the size of the matrix $M$, which increases factorially with the number of particles. 
Our previous work in \cite{walker2019transient} showed a collection of approximately 380 beads condensing into clusters dynamic clusters of approximately 5-10 beads in suitable parameter regimes. In future work, we would like to apply our methods to larger numbers of beads using forces directly taken from biological modeling. The largest challenge to this will be addressing the state space of the switching process, which increases exponentially with the number of beads.  
The algorithm does not use the matrix $M(x,p)$ in full, only the derived (scalar) Hamiltonian $\mathcal{H}(x,p)$ arising from the greatest eigenvalue. Since $x$ and $p$ only scale up in dimension linearly with number of particles, this suggests an alternative approach in which we construct an approximation either for the map $(x,p) \to \mathcal{H}(x,p)$, which is used to implicitly solve for $p$, or even directly for the map $x \to p$ (noting that the direction $\frac{d\phi}{ds}$ of the string would also need to be included). Recently machine learning methods have been used to develop approximations to challenging-to-compute mathematical functions, and so this is an approach we would like to incorporate into our method in future work.

The WKB ansatz approach also allows for the computation of additional terms by following through the asymptotic expansion to higher order. In particular, the next term in the series would give the full pre-exponential term that would allow for computation of the intercept of the escape time asymptotics as shown in \cref{fig:res_1d,fig:string_mc}. While the current work focused on developing the machinery to compute most-probable transition paths under our framework, future work could expand on this and compute this pre-exponential coefficient.

\section*{Acknowledgements}

We would like to thank Jay Newby for insightful conversations and advice on the computation of quasipotentials for switching systems. This work was partially supported by the National Science Foundation under grants DMS-1816630 and DMS-1816394.

\appendix

\section{Monte Carlo Simulation}
\label{sec:meth_monte_carlo}
To validate our quasipotential, we compare it to Monte Carlo simulations of the system. These simulations are taken by numerically evaluating \cref{eqn:main_sde} using a modified Euler-Maruyama method, to incorporate both the SDE and the stochastic switching. We track the current switching state $s$ in addition to the position $x$, and update the position in each step as 
\begin{equation}
\label{eqn:euler_maruyama}
    x_{\textrm{new}} = x + \Delta t F(x; s) + \sqrt{2 \epsilon} \Delta B
\end{equation}
where $F(x;s)$ is the drift function at position $x$ in state $s$, and $\Delta B \sim \mathcal{N}(0, \Delta t)$ is the increment of Brownian noise. Then, we compute any changes to the state variable $s$. The specifics of this update depend on the nature of the switching, but in general involve checking whether any changes occur within the timestep $\Delta t$ by drawing waiting times from exponential distributions. Due to the memoryless property of exponential random variables, we may safely redraw in future time steps any waiting times that do not correspond to a transition within the current timestep, in case the transition rate changes between timesteps. 
In this work, we make a first-order approximation and treat all switches as occurring at the beginning of the timestep.

To empirically estimate the mean escape time associated with a particular $\epsilon$, we initialize a Monte Carlo simulation in a state corresponding to a stable equilibrium of the deterministic dynamics, as obtained through descent of the deterministic force.
We proceed to simulate \cref{eqn:euler_maruyama} until a termination condition is reached indicating escape from the basin of attraction that the simulation began in. The escape time is then recorded as $\tau_k$. The simulation may also reach maximum time $T$ without exiting the original basin of attraction, in which we take $\tau_k = T$. We assume escape times follow an exponential distribution with mean $\mu$ and use the maximum likelihood estimator (MLE) for the mean of an exponential distribution given samples capped at a maximum time, given by~\footnote{\url{https://www.itl.nist.gov/div898/handbook/apr/section4/apr412.htm}}
\begin{equation}
    \hat{\mu} = \frac{\sum_k{\tau_k}}{\sum_k I\left(\tau_k < T\right)}
\end{equation}
(Note that as $T \to \infty$, this reduces to the simple mean of the samples $\tau_k$.

We then repeat the above process for a sequence of values of $\epsilon$, computing a relationship $\hat{\mu}(\epsilon)$.
We expect a linear relationship between $\frac{1}{\epsilon}$ and $\log{\hat{\mu}}$
\begin{equation}
    \log \hat{\mu} = a + \frac{b}{\epsilon}
\end{equation}
where the value $b$ is corresponds theoretically with the quasipotential barrier height, and $a$ would be given by the higher-order pre-exponential term.

\section{Derivation Details}
\subsection{Derivation of the Hamiltonian}
\label{sec:hamiltonian_derivation}
Here we will present the computation arising from plugging in the WKB ansatz~\cref{eqn:wkb_ansatz}
\begin{equation*}
    p_s(x) = r_s \exp\left( -\frac{1}{\epsilon} W(x)  \right)
\end{equation*}
to the steady state equation~\cref{eqn:steady_state}
\begin{equation*}
    0 = -\sum_i \frac{\partial}{\partial x_i}\left[v^s_i p_s\right] + \epsilon \sum_i \frac{\partial^2}{\partial x_i^2} \left[ p_s \right] + \frac{1}{\epsilon} \left[ S {p} \right]_s
\end{equation*} (both reproduced for clarity)
and collecting lowest order terms in $\epsilon$, which will be the $\frac{1}{\epsilon}$ terms.

The spatial derivatives of the WKB ansatz are given by
\begin{equation*}
    \frac{\partial}{\partial x_i} \exp\left(-\frac{W(x)}{\epsilon} \right) = \exp\left(-\frac{W(x)}{\epsilon} \right) \left[ -\frac{1}{\epsilon} \frac{\partial W}{\partial x_i}\right]
\end{equation*}
and
\begin{equation*}\begin{aligned}
    \frac{\partial^2}{\partial x_i^2} \exp\left(-\frac{W(x)}{\epsilon} \right) = \exp\left(-\frac{W(x)}{\epsilon} \right) \left[ -\frac{1}{\epsilon} \frac{\partial W}{\partial^2 x_i^2}\right] \\
    + \exp\left(-\frac{W(x)}{\epsilon} \right) \left[ \frac{1}{\epsilon} \frac{\partial W}{\partial x_i}\right]^2.
\end{aligned}\end{equation*}
For the drift term, we obtain a $\frac{1}{\epsilon}$ term by differentiating once the exponential, thus giving us 
\begin{equation*}
    -\sum_i \frac{\partial}{\partial x_i}\left[v^s_i p_s\right] \sim \frac{1}{\epsilon} r_s \exp\left(-\frac{W(x)}{\epsilon} \right) \sum_i v_i^s \frac{\partial W}{\partial x_i}.
\end{equation*}
For the diffusion term, we obtain a $\frac{1}{\epsilon}$ term by differentiating the exponential twice and combining with the $\epsilon$ pre-factor, giving us 
\begin{equation*}
    \epsilon \sum_i \frac{\partial^2}{\partial x_i^2} \left[ p_s \right] \sim \frac{1}{\epsilon}  r_s \exp\left(-\frac{W(x)}{\epsilon} \right) \sum_i\left(\frac{\partial W}{\partial x_i}\right)^2.
\end{equation*}
Finally,  the switching term itself is a $\frac{1}{\epsilon}$ term, given by
\begin{equation*}
    \frac{1}{\epsilon} \left[ S {p} \right]_s = \frac{1}{\epsilon} \exp\left( -\frac{W(x)}{\epsilon}   \right) \left[ S {r} \right]_s.
\end{equation*}
Together, \cref{eqn:steady_state} reduces to, at lowest order in $\epsilon$, 
\begin{equation}\begin{aligned}
    0 \hspace{0.1cm} = \hspace{0.2cm}  &\frac{1}{\epsilon} r_s \exp\left( -\frac{W(x)}{\epsilon}   \right) \sum_i v_i^s \frac{\partial W}{\partial x_i} \\+ &\frac{1}{\epsilon}  r_s \exp\left( -\frac{W(x)}{\epsilon}   \right) \sum_i\left(\frac{\partial W}{\partial x_i}\right)^2 \\+ &\frac{1}{\epsilon} \exp\left( -\frac{W(x)}{\epsilon}   \right) \left[ S {r} \right]_s.
\end{aligned}\end{equation}

We can cancel out the common term of $\frac{1}{\epsilon} \exp\left( -\frac{W(x)}{\epsilon}  \right)$, and then rewrite the resulting equation as a matrix equation by defining vector $r$ with components $r_s, s=1\dots n$  and turning the advection (drift), diffusion, and switching terms into the matrices A, D, and S respectively: 
\begin{equation}
    \left[A + D + S \right] {r} = M {r} = {0}
\end{equation}
with the diffusion matrix $D$ defined as 
\begin{equation}\label{eq:dfn_D}
    D = \left( \sum_i \left(\frac{\partial W}{\partial x_i}\right)^2 \right) \mathbb{I} 
\end{equation}
where $\mathbb{I}$ represents the identity matrix, and
 the advection matrix $A$ defined as 
\begin{equation}
\label{eqn:dfn_A}
    A = \textrm{diag} \left( V \nabla W \right)
\end{equation}
(note: diag here indicates the mapping of a vector to the square matrix with it on the diagonal),
and the matrix $S$ unchanged from its original definition. Note also that $D$ and $A$ are diagonal matrices, so the only off-diagonal contributions come from $S$, which is not dependent on $W$.

\subsection{Derivatives of the Hamiltonian}
\label{sec:eig}
In order to apply Newton's method for finding critical points, we will need to compute the gradient and Hessian of $\mathcal{H}(x,p)$ with respect to the momentum variable $p$. 
To do so, we will use the formulas of \cite{magnus1985differentiating} in differentiating eigenvalues of a matrix with respect to the entries of that matrix. Consider a real, square matrix $M=M_0+dM$ such that $M_0 u_0 = \lambda_0 u_0$ and $v_0 M_0$ = $\lambda_0 v_0$, and then consider the function $\lambda(dM)$ s.t. $\lambda(0)=\lambda_0$. Using the superscript $+$ to refer to the Moore-Penrose inverse, we have
\begin{align*}
    d\lambda &= \frac{v_0^T dM u_0}{v_0^T u_0} \\
    d^2\lambda &= \frac{2 v_0^T (dM) K_0(\lambda_0 I - M_0)^+ K_0 (dM) u_0}{v_0^Tu_0} \\
    K_0 &= I - \frac{u_0v_0^T}{v_0^Tu_0}
\end{align*}
We now consider the case that $M$ is in fact a function of the momentum variable ${p}$, centered at a value ${p}_0$. In this case we have $M({p}_0) = M_0$. Applying the chain rule, we get the formulas
\begin{align}
    \frac{d\lambda}{dp_i} = & \frac{v_0^T M_{p_i} u_0}{v_0^T u_0} \label{eqn:eig_drv_1}\\
    \frac{d^2 \lambda}{dp_i dp_j} = & \frac{2 v_0^T M_{p_i} K_0(\lambda_0 I - M_0)^+ K_0 M_{p_j} u_0}{v_0^Tu_0} \\&+ \delta_{ij}\frac{v_0^T M_{p_i,p_i} u_0}{v_0^T u_0} \label{eqn:eig_drv_2}
\end{align}
where $p_i=\partial W/\partial x_i$ for $i=1\dots m$.

Because $\mathcal{H}(x,p)$ is simply the largest eigenvalue of the matrix $M$, we can compute the gradient and Hessian $\mathcal{H}_p$ and $\mathcal{H}_{pp}$ with direct application of \cref{eqn:eig_drv_1,eqn:eig_drv_2}.

\subsection{Relationship between deterministic dynamics and quasipotential}
\label{sec:det_fp}
In this section we provide a proof that fixed points $x^d$ of the deterministic dynamics correspond to points where $\nabla_p \mathcal{H}(x^d, 0) = 0$, showing that $\nabla W=0$ corresponds to a minimum and therefore unique solution of $\mathcal{H}(x^d, \nabla W)=0$.

	Let ${x}^d$ be a fixed point of the deterministic dynamics such that $F_{\textrm{det}}({x}^d) = 0$, and let ${r}^d$ be the probability vector across the states,
	and let the force term for each switching state $v_i^s$ be represented in a matrix $V_{is} = v^s_i$
	such that
    \begin{align*}
        S(x^d) r^d &= {0} && \textrm{ because }r^d \textrm{ is a nullvector} \\
         V(x^d) r^d &= {0} && \textrm{ because } x^d \textrm{ is a fixed point}
    \end{align*}
   Recall the definition $M(x, \nabla W)=A(x, \nabla W) + D(\nabla W) + S(x)$. One can see from the definitions in \cref{eq:dfn_D,eqn:dfn_A} that $A$ and $D$ vanish when $\nabla W=0$, so that $M(x, 0) = S(x)$. It follows that $M(x, 0)r = S(x) r = 0$, so $0$ is an eigenvalue of $M(x,0)$ and therefore $\mathcal{H}(x,0)=0$ for any $x$.
    
    We apply \cref{eqn:eig_drv_1} to compute $\frac{d\mathcal{H}}{dp_i}$, which we will show to be $0$ for all $i$. 
    \begin{equation*}
        \frac{d\mathcal{H}}{dp_i} = \frac{v_0^T (dM/dp_i) u_0}{v_0^T u_0}
    \end{equation*}
    where $u_0$, $v_0$ are the right and left eigenvectors, respectively, of $M(x_d, 0) = S(x_d)$, corresponding to an eigenvalue $\lambda = 0$. Recalling that $S$ is a CTMC transition rate matrix, the right eigenvector $u_0$ will naturally be the steady-state vector $r^d$. Similarly, as a transition rate matrix its columns always sum to $0$, and so the ones vector will be a left eigenvector with eigenvalue $0$, and thus $v_0 = [1, 1, 1 \dots 1 ]^T$. 
    
    Noting that 
    \begin{align*}
        \frac{dA}{dp_i} &= \textrm{diag}(V_i) \\
        \frac{dD}{dp_i} &= 2 p_i I = 0 \quad (p_i=0)
    \end{align*}
    we get that 
    \begin{equation*}
        dM(x^d, 0) u_0 =  \textrm{diag}(V_i) r^d = v_0^T V_i^T r^d = \frac{d}{dt} x^d = 0
    \end{equation*}
    as $x_d$ is a fixed point of the deterministic dynamics.

\section{Algorithm Implementation Details}

Note that full code for reproducing all results is available on Github at \url{https://github.com/bwalker1/quasi-string-reprod}.

\subsection{Quasipotential Implicit Solver}
\label{sec:app_quasi_solver}

The core Newton update from \cite{newby2014spontaneous} are given by the iterative update equations
\begin{align}
p_{n+1} &= p_n + \mathcal{H}_{pp}^{-1}\left[\lambda_n \frac{d\phi}{ds} - \mathcal{H}_p\right] \label{eq:piterate}\\
\lambda_n &= \sqrt{\frac{\mathcal{H}_p \mathcal{H}_{pp}^{-1} \mathcal{H}_p - 2\mathcal{H}}{d\phi^T \mathcal{H}_{pp}^{-1} d\phi}} \label{eq:lambiterate}
\end{align}
starting from an initial guess for $p_0$,
where $\mathcal{H}, \mathcal{H}_p, \mathcal{H}_{pp}$ are evaluated at $x, p_n$, and the derivatives w.r.t $p$ are computed  in \cref{sec:eig}. If the starting guess is sufficiently close to the correct value, \cref{eq:piterate}  will converge to the value $p$ that satisfies \cref{eqn:cond1,eqn:cond2}, which is the value of $\nabla W$. However, in practice, we find that for our problems the Newton method frequently fails to converge, and so we add a modification based on the structure of the optimization problem to guarantee convergence.

The solution to \cref{eqn:cond2} under the constraint \cref{eqn:cond1} can be equivalently posed as the unique solution of
\begin{equation}
\textrm{arg}\max_p \left[ p \cdot \frac{d\phi}{ds} \right] \qquad \textrm{s.t. } \mathcal{H}(x,p)=0
\end{equation}
as the surface $\mathcal{H}(x,p)=0$ is convex.
In other words, \cref{eqn:cond2} can be viewed as a maximization problem. Correspondingly, we require that each iteration increase the objective quantity $p \cdot \frac{d\phi}{ds}$. If the Newton's step fails to do so, we move a small distance $\kappa$ in the direction of $\frac{d\phi}{ds}$ projected onto the normal to the $\mathcal{H}(x,p)=0$ surface:
\begin{align*}
p_n^* = p_n + \kappa \, \textrm{proj}_{\bot \mathcal{H}_p}\left( \frac{d\phi}{ds} \right).
\end{align*}
We then use the root-finding version of Newton's method to find a $p_{n+1}$ such that $\mathcal{H}(x, p_{n+1})=0$, starting from $p_n^*$, until convergence: 
\begin{equation}
p_n^* = p_n^* - \frac{\mathcal{H}(x, p_n^*)}{\mathcal{H}_p(x,p_n^*)}.
\end{equation}
If this did not increase the objective quantity $p \cdot \frac{d\phi}{ds}$, we reduce $\kappa$ by a factor of 2 and try again. Because the surface is convex and we are moving in the objective direction projected onto the surface normal, we are guaranteed an improvement for sufficiently small $\kappa$. However, it is desireable that $\kappa$ not be too much smaller than necessary to reduce the number of iterations until Newton's method begins to converge.

We find that with this modification we always reach a point where Newton's method begins to converge quadratically to the true solution.

\subsection{String Method}
\label{sec:app_string}

We apply the climbing string method as described in \cite{ren2013climbing}. The only modification is the replacement of the gradient of the potential $\nabla U$ with the gradient of the quasipotential $\nabla W$. These differ in that whereas $\nabla U$ evaluated at an image $\phi_i$ would depend only on the value of $\phi_i$, $\nabla W$ depends on both $\phi_i$ and the direction of the string $\frac{d\phi}{ds}|_{\phi=\phi_i}$. We estimate this using a centered finite difference approximation, 
\begin{equation}
    \frac{d\phi}{ds}|_{\phi=\phi_i} \approx \frac{\phi_{i+1} - \phi_{i-1}}{2h}
\end{equation}
which turns \cref{eqn:cond2} into the numerical
\begin{equation}
\left. \frac{d\phi}{ds}\right|_{\phi=\phi_i} \parallel \phi_{i+1} - \phi_{i-1}.
\end{equation}
Note that we drop the denominator $\frac{1}{2h}$ as it does not affect the direction and therefore is irrelevant to the parallel condition. This however means that the update order affects the algorithm. In our case, we compute the directions $\frac{d\phi}{ds}$ at all images first, and then perform updates simultaneously.

%

\end{document}